\documentclass[aps,prb,preprint,showpacs,groupedaddress]{revtex4-1}

\usepackage{graphicx}
\usepackage{bm}
\usepackage{amsmath}

\begin{document}


\title{Dynamics of twisted vortex bundles and laminar propagation of the vortex front }


\author{E. B. Sonin}
\affiliation{Racah Institute of Physics, Hebrew University of
Jerusalem, Jerusalem 91904, Israel}


\date{\today}

\newcommand{\be}{\begin{equation}}
\newcommand{\ee}{\end{equation}}
\newcommand{\el}[1]{\label{#1}\end{equation}}
\newcommand{\bem}{\begin{eqnarray}}
\newcommand{\eem}{\end{eqnarray}}
\newcommand{\eml}[1]{\label{#1}\end{eqnarray}}
\newcommand{\eq}[1]{Eq.~(\ref{#1})}
\newcommand{\Eq}[1]{Equation~(\ref{#1})}

\begin{abstract}
The paper is studying the dynamics of twisted vortex bundles, which were detected in experimental investigations of superfluid turbulence in superfluid $^3$He-B. The analysis shows that a linear torsion oscillation of a vortex bundle  is a particular case of the slow vortex mode related with the inertial wave, which was already investigated in the past in connection with observation of the Tkachenko waves in superfluid $^4$He and the experiments on the slow vortex relaxation in superfluid $^3$He-B.  The paper addresses also a twisted vortex bundle terminating at a lateral wall of a container starting from the elementary case when the bundle reduces to a single vortex.  The theory  considers the laminar regime of the vortex-bundle evolution and investigates the Glaberson--Johnson--Ostermeier instability of the laminar regime, which is a precursor for the transition to the turbulent regime at strong twist of the bundle.
 The propagation and the rotation velocities of the vortex front (the segment of the vortex bundle diverging to the wall) can be found from the equations of balance for the linear and the angular momenta, and the energy. It is demonstrated that the vortex front can move with finite velocity even in the absence of mutual friction (the $T=0$ limit). The theory is compared with experimental results  on vortex-front  propagation in superfluid $^3$He-B.
\end{abstract}

\pacs{}

\pacs{67.30.hb,47.15.ki, 67.30.he}
\maketitle



%

\section{Introduction}

Theoretical and experimental investigations of various vortex structures  in classical and superfluid hydrodynamics have a long history \cite{Saf,D}.   In particular,  helix vortices, which  appear in wakes of propellers and other spinning bodies, were intensively studied in classical hydrodynamics \citep{Oku}. Recently, they became an object of investigation also in superfluid hydrodynamics, however, not as individual single vortices, but as bundles of helix vortices, also called {\em twisted vortex bundles}. The interest to twisted vortex bundles arose in connection with  experimental studies of  the transient process of establishing of stable vorticity in rotating superfluid $^3$He-B \cite{Twist}. The twisted vortex bundle appears in an originally vortex-free rotating container with a superfluid  when vortex lines being injected at container's bottom expand into the rest part of the container.  
 The experimental and theoretical studies of the twisted vortex bundle addressed possible turbulence of the vortex front and the transition from the laminar to the turbulent regime\cite{Finne,VolRe,Elt07,Elt10,EltProg,Elt11}. Meanwhile a proper dynamical theory of the laminar regime of the propagating vortex front is still lacking. There were estimations of the connection between the dissipation rate and the velocity of the vortex front, but the front velocity was not derived from the vortex dynamics. However, without a satisfactory theory of the laminar regime it is impossible to have a reliable physical picture of the more complicated turbulent regime or of  the transition between the laminar and the turbulent regime. The first step in this direction was undertaken in Ref.~\onlinecite{SN}, but for the simplest case when the vortex bundle is not twisted and the force driving the vortex front is very weak.
The goal of the present work is to suggest a more realistic and general theory.

The paper starts from the overview of the uniformly twisted vortex bundle in Sec.~\ref{UnTwist}. This is a generalization of the previous analysis  \cite{Twist} on the case when the vortex-line tension is important.  The section presents the analysis of the Glaberson--Johnson--Ostermeier instability \cite{glabL,glab} of the twisted vortex bundle (Sec.~\ref{GlabIn}), which demonstrates that the laminar regime becomes unstable at rather weak twist. The instability can be considered as a precursor of the transition to the turbulent regime.
Section~\ref{TM} analyzes linear oscillations of the vertically uniform vortex  bundle on the basis of the general theory of slow vortex oscillations developed decades ago \cite{RMP}. Section~\ref{front} presents  the dynamical theory of a vortex  bundle terminating at a lateral wall starting from the elementary case when the vortex bundle reduces to a single vortex.  The balance equation for the linear and the angular momenta, and the energy are derived and then used for calculation of the vortex front velocities of propagation along and rotation around the container axis.   A special attention is devoted to the limit of vanishing mutual friction (the $T=0$ limit) and to qualitative comparison with the experiment. Concluding discussion is presented in the last Sec.~\ref{concl}.

\begin{figure}[b]
 \includegraphics[width=0.85\linewidth]{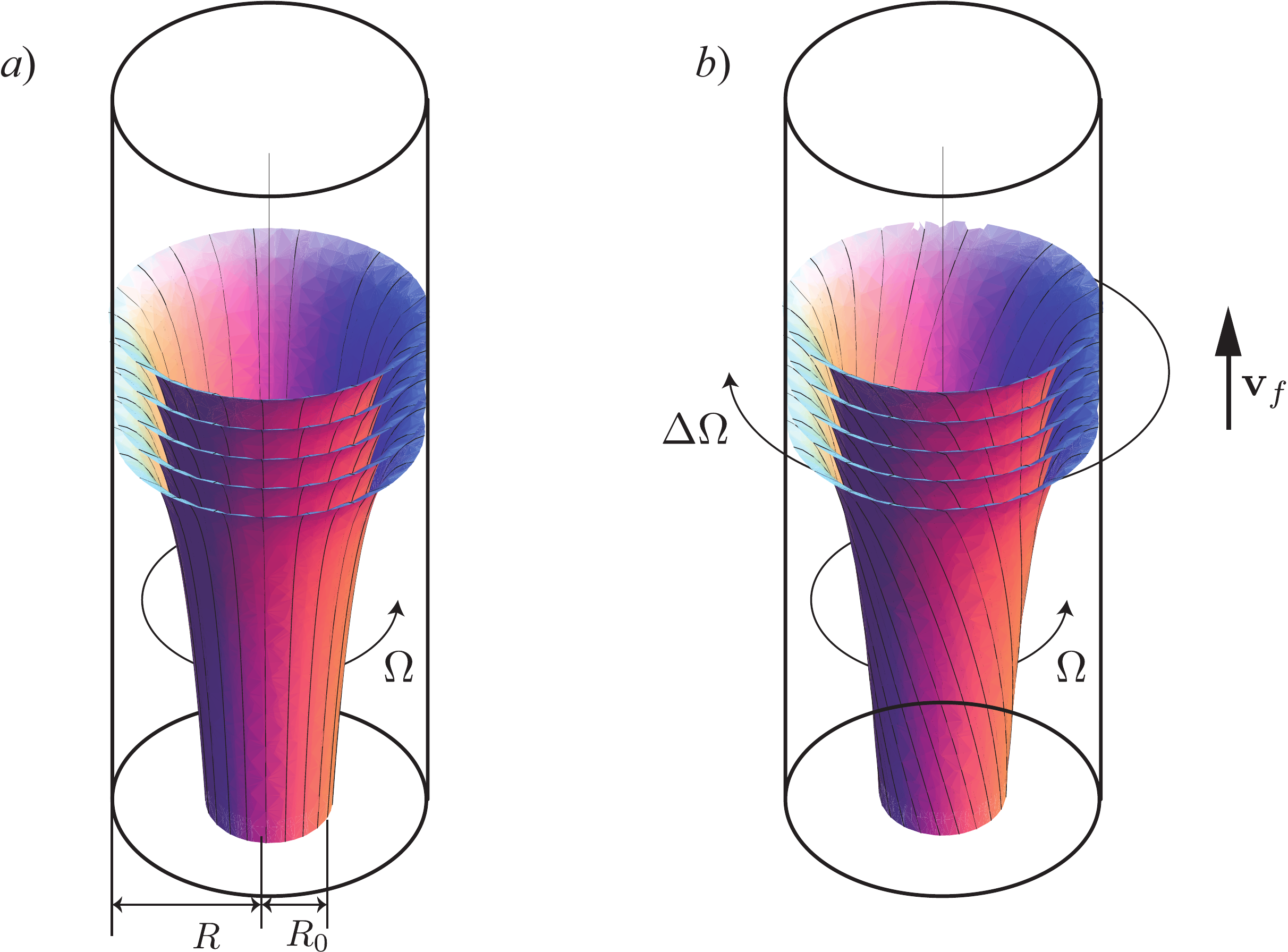}%
 \caption{Untwisted and twisted vortex bundle. (a) Untwisted vortex bundle. Any vortex line is in an axial plane, i.e., is not twisted. Such a structure corresponds to the equilibrium state in the coordinate frame rotating with the angular velocity $\Omega$ (Ref.~\onlinecite{SN}). (b) Twisted vortex bundle. The twisted stem of the bundle is at rest   in the coordinate frame rotating with the angular velocity $\Omega$. The vortex front (whorl) propagates with the velocity $v_f$ along the axis $z$ and rotates around the axis $z$ with the relative velocity $\Delta \Omega =\Omega_f-\Omega $ with respect to the container, the normal fluid, and the bundle stem rotating together.  }  \label{f1}
 \end{figure}

\section{Uniformly twisted stationary vortex bundle} \label{UnTwist}

\subsection{Continuous vorticity} \label{CV}

We start from the purely continuum approach to superfluid vorticity neglecting quantum elastic effects (line tension of vortex lines and Tkachenko shear rigidity), which are proportional to the circulation quantum $\kappa$.  So vorticity inside an untwisted vortex bundle of radius $R_0$ in a container of radius $R$ [Fig.~\ref{f1}(a)] is continuous and uniform,  the azimuthal superfluid  velocity field being 
\begin{equation}
v_{s\phi}=\left\{\begin{array} {cc} \Omega r & r<R_0, \\
{\Omega R_0^2 \over r}  & R>r>R_0 \end{array} \right.   .\label{clust}
\end{equation}

In a uniformly twisted vortex bundle [Fig.~\ref{f1}(b)] any vortex line forms a helix of  pitch $2\pi /Q$  moving along a cylinder coaxial with the $z$ axis, where $Q = \omega_{0\phi}/ \omega_{0z}r=u_{\phi}/ u_{z}r$ is the wavenumber of the twist (called later {\em twist}), $ \omega_{0\phi}$ and  $\omega_{0z}$ 
are the azimuthal and the axial components of the vorticity $\bm \omega_0 =\bm \nabla \times \bm v_s$,  $\bm v_s$ is the superfluid velocity, and $u_\phi$ and $ u_z$ are the azimuthal and the axial components of the  displacements $\bm u$ of vortex lines from positions in the untwisted bundle described by \eq{clust}. In an untwisted bundle  the vorticity $\bm \omega_0$ has the only axial component $\omega_0=\omega_{0z}=2\Omega$. In a twisted bundle azimuthal components $\omega_{0\phi}$ and $u_\phi$ appear, but radial components $\omega_{0r}$ and $u_r$  still vanish. In the continuum approach the twist $Q(r)$ can be an arbitrary function of the distance $r$ from the axis. This is a consequence of ignoring  Tkachenko shear rigidity of the crystalline vortex bundle: shear deformation of the bundle does not cost any energy. On the other hand, choosing a $r$-independent twist $Q$ one may expect that Tkachenko rigidity will not have an essential effect on the twisted bundle. Later it will be shown (Sec.~\ref{Tk}) that this expectation is true only for a stationary twist, while considering the torsional oscillation of the bundle the role of Tkachenko rigidity is important. 

We are looking for the stationary twisted state in the coordinate frame rotating with the angular velocity $\Omega$. In the continuum theory the vortex line moves with the velocity $\bm v_L$ coinciding with average superfluid velocity $\bm v_s$ (Helmholtz's theorem), and the vortex lines are at rest if the superfluid velocity in the rotating frame has no component normal to the vortex line. This yields the condition \cite{Twist}
\be
[v_{s\phi}(r) -\Omega r] \omega_{0z}(r)-v_{sz}(r)  \omega_{0\phi}(r)=0.
  \el{stat}
One can check by substitution that the following pattern satisfies this condition:
\bem
v_{s\phi}={(\Omega+Qv_0)r\over 1+Q^2r^2},~~v_{sz}={v_0-\Omega Qr^2\over 1+Q^2r^2}, 
\nonumber \\
\omega_{0z}=-{1\over r}{\partial(r v_{s\phi} )\over \partial r}={2(\Omega+Qv_0)\over (1+Q^2r^2)^2},~~
\omega_{0\phi}={\partial  v_{sz}\over \partial r} ={2Qr(\Omega+Qv_0)\over (1+Q^2r^2)^2},
\eml{Thun}
where $v_0$ is a constant determined by a natural condition that the total axial mass flow $2\pi\int_0^R v_{sz}(r)r\,dr$ must vanish in a closed container. In the vortex-free region $R>r>R_0$ the axial velocity $v_{sz}$ does not depend on $r$ and if $R \gg R_0$ $v_{sz}$ must be very small. This means that the absence of the axial mass flow requires that $ v_{sz}(R_0)=0$ and correspondingly $v_0=Q\Omega R_0^2$.  In the opposite limit $R_0 \approx R$ the bundle occupies the whole cross-section of the container, and the axial mass flow vanishes if
\bem
v_0 ={\Omega \over Q}\left[\frac {Q^2 R^2}{\ln(1+Q^2 R^2)} -1\right] .
   \eml{v0} 
Then the velocity field in the twisted bundle is
\bem 
v_{s\phi}={\Omega r\over 1+Q^2r^2} \frac {Q^2 R^2}{\ln(1+Q^2 R^2)} ,
~~
v_{sz} ={\Omega \over Q}\left[\frac {Q^2 R^2}{(1+Q^2r^2)\ln(1+Q^2 R^2)} -1\right].
   \eem    
   Knowing this field one can find the energy $e$ and the angular momentum $m_z$  per unit length of the bundle. For the bundle filling the whole container ($R\approx R_0$)  they are given by 
\bem
e =2\pi\rho_s\int_0^R {v_{s\phi}^2+v_{sz}^2\over 2}r\,dr    ={\pi \rho_s\Omega ^2R^2\over 2}  \left[\frac{R^2}{ \ln(1+Q^2 R^2)}-{1\over Q^2}\right], 
  \eem 
\bem
m_z=2\pi\rho_s\int_0^R v_{s\phi} r^2\,dr= \pi \rho_s \Omega R^2  \left[\frac{R^2}{ \ln(1+Q^2 R^2)}-{1\over Q^2}\right].
  \eem 
Twisting of the bundle leads to the flux of the angular moment along the vertical axis. Using the known hydrodynamic expression
\be
\Pi_{ij}=P\delta_{ij}+ \rho_s v_{si} v_{sj}
   \el{flux} 
for the momentum flux tensor of the superfluid component, the flux of the angular momentum along the axis $z$ is given by
\be
J_m=2\pi\rho_s\int_0^R v_{s\phi} v_{sz} r^2\,dr  
=- {\pi \rho_s\Omega ^2 R^2\over Q^3}\left[\frac{Q^4 R^4}{ \ln^2(1+Q^2 R^2)(1+Q^2 R^2)}-1\right]. 
 \ee

The proper thermodynamic potential for a liquid in a   container rotating with constant angular velocity $\Omega$ is the Gibbs potential  $g=e-\Omega m_z$.  An important property of the continuous-vorticity  approach is that the Gibbs potential of the uniform vortex bundle is always negative independently from how strong the twist $Q$ is.  
But in the presence of quantum-vorticity effects, namely, line tension,  the condition $g=0$ can be satisfied as shown in the next subsection. 
This important for dynamics of the vortex-front propagation (Sec.~\ref{front}). 

\subsection{Quantum-vorticity effect: vortex-line tension} \label{QLT}

Now we generalize the analysis of the twisted vortex bundle by taking into account vortex-line tension. The widely accepted method of dealing with the line-tension effects is the Hall--Vinen--Bekarevich--Khalatnikov (HVBK)  theory \cite{Khalatnikov2000,RMP}.  This theory uses the local-induction approximation, when the contribution of line tension  to the energy and the superfluid velocity depends on the circulation quantum and the curvature radius of the vortex lines. Helmholtz's theorem is still valid but now  the vortex velocity  $\bm v_L$ coincides with the local superfluid velocity $\bm v_{sl}$, which differs from the 
average superfluid velocity $\bm v_s$ and in the local-induction approximation is given by
\be 
\bm v_{sl} = \bm v_s +\nu_s\bm \nabla \times \hat s,
         \ee 
where $\hat  s=\bm \omega_0/\omega_0$ is the unit vector tangent to the vortex lines,    $\nu_s=(\kappa/4\pi) \ln (r_v/r_c)$ is the line-tension parameter, and $r_v$ and $r_c$ are the intervortex distance and the core radius respectively.
Then Eq.~(\ref{stat}), which provides the condition that vortex lines are at the rest in the rotating coordinate frame, transforms to
\be
\left[v_{s\phi}(r) -\Omega r+{\nu_sQ^2 r \over (1+Q^2r^2)^{3/2}} \right] \omega_{0z}(r)-\left[v_{sz}(r) +{\nu_sQ(2+Q^2 r^2) \over (1+Q^2r^2)^{3/2}}\right] \omega_{0\phi}(r)=0.
  \el{statLT}
 The superfluid velocity and the vorticity  fields satisfying this condition are [cf.  \eq{Thun}]
 \bem
v_{s\phi}={(\Omega+Qv_0)r\over 1+Q^2r^2}+\frac{\nu_sQ^2 r}{ (1+Q^2r^2)^{3/2}},~~v_{sz}={v_0-\Omega Qr^2\over 1+Q^2r^2}- \nu_sQ^3 {r^2\over (1+Q^2r^2)^{3/2}}, 
\nonumber \\
\omega_{0z}={2(\Omega+Qv_0)\over (1+Q^2r^2)^2}+\frac{2\nu_sQ^2(1-Q^2r^2/2)}{ (1+Q^2r^2)^{5/2}},~~
\omega_{0\phi} ={2Qr(\Omega+Qv_0)\over (1+Q^2r^2)^2}+\frac{\nu_sQ^3r(2-Q^2r^2)}{ (1+Q^2r^2)^{5/2}}.
        \eml{ThunLT}
Further in this subsection only the case $R \approx R_0$ is considered. The axial mass flow vanishes if
\bem
v_0 ={\Omega \over Q}\left[\frac {Q^2 R^2}{\ln(1+Q^2 R^2)} -1\right] +{2\nu_sQ\over \ln (1+Q^2R^2)}\left({2+Q^2R^2\over \sqrt {1+Q^2R^2}}-2 \right).
   \eem 

 The density of the Gibbs potential $g=e-\Omega m_z$ includes the vortex line-tension term proportional to $\nu_s$:
\bem
g=2\pi\rho_s\int_0^R {v_{s\phi}^2+v_{sz}^2\over 2}r\,dr  +2\pi \rho_s \nu_s \int_0^R  \omega_0r\,dr  -2\pi \Omega \rho_s\int_0^R v_{s\phi} r^2\,dr.
    \eem 
Substituting  the superfluid velocity field given by \eq{ThunLT} and bearing in mind that the modulus of vorticity vector is $\omega_0=\omega_{0z}\sqrt{1+Q^2R^2}$ one obtains that 
\bem
g=\pi \rho_sR^2 {v_0^2\over 2} -\frac{\pi \rho_s(\Omega +Qv_0)^2}{2Q^4}[Q^2 R^2-\ln (1+Q^2R^2)]
\nonumber \\
+ 4\pi \rho_s\nu_s{\Omega +Qv_0\over Q^2}\left(1-\frac{1}{(1+Q^2R^2)^{1/2}}\right)+\frac{\pi\rho_s \nu_s^2  }{ 2}\left[{5 Q^2R^2 \over 1+Q^2R^2}-\ln (1+Q^2R^2)\right]. 
   \eem   
Line tension contributes not only to the energy, but also to the momentum flux tensor, which becomes \cite{Khalatnikov2000,RMP}
\be
\Pi_{ij}=P\delta_{ij}+ \rho_s v_{si} v_{sj} +\rho_s \nu_s\left(\omega_0\delta_{ij}-{\omega_{0i} \omega_{0j}\over \omega_0} \right).
   \ee 
Using this expression one can find the axial flux of the angular momentum along the axis $z$:
\be
J_m=2\pi\rho_s\int_0^R \Pi_{\phi z}r^2\,dr =2\pi\rho_s\int_0^R\left( v_{s\phi} v_{sz} -\nu_s{\omega_{0\phi} \omega_{0z} \over \omega_0 } \right)r^2\,dr. 
 \ee 
After substitution of the values of the components of the velocity $\bm v_s$ and the vorticity $\bm \omega_0$ and following integration over the cross section of the cylindric container one obtains 
\bem
J_m
= -  \frac{\pi\Omega^2 R^2}{ Q^3}\left[{Q^4 R^4\over  (1+Q^2R^2)  \ln^2 (1+Q^2R^2)}-1\right]
-{4\pi \Omega \tilde \nu_sQ R^4\over  (1+Q^2R^2) \ln^2 (1+Q^2R^2) }
\nonumber \\
- \frac{4\pi \tilde \nu_s^2 QR^2}{(1+Q^2R^2)\ln ^2(1+Q^2R^2)} 
-{\pi \nu_s^2 Q^3R^4\over (1+Q^2R^2)^2},
           \eem 
where
\be
\tilde \nu_s=\nu_s\left({2+Q^2R^2\over \sqrt {1+Q^2R^2}}-2 \right)
  \ee 
is the renormalized line-tension parameter taking into account the bundle twist $Q$. The expression for the flux of the angular momentum along the axis $z$ can be also derived from the thermodynamic definition $\left.  \left. J_m=-\partial e/\partial \nabla_z \varphi \right|_{m_z}=-\partial g/\partial \nabla_z \varphi \right |_{\Omega}$, where subscripts mean that derivatives with respect to $Q=\nabla_z \varphi$ are taken at fixed $m_z$ or $\Omega$ respectively. But this method is complicated by the condition that the derivatives must be calculated at constant number of vortices, whereas derivation of the  flux of the angular momentum from the hydrodynamic momentum-flux tensor is more straightforward. 

The line tension terms $\propto \nu_s$ make possible the condition $g=0$, which provides the balance of forces on the vortex front separating the vortex bundle from the vortex-free region. In the limit $QR \to  0$ the condition  $g=0$ determines the angular velocity $\Omega = 8 \nu_s/R^2$, at which the vortex bundle terminated at the lateral wall is in the equilibrium with the rotating container. This velocity was found in Ref. \onlinecite{SN}. In the opposite limit   $QR \to  \infty$ the condition is satisfied at $\Omega = 2 \nu_sQ/R$. So the strong twist $Q$ can facilitate the balance of forces 
 on the vortex front and make possible the stationary motion of the vortex front without friction (see Sec.~\ref{front}).

\subsection{Stability of twisted vortex bundle} \label{GlabIn}

An important feature of a twisted vortex bundle is a mass flow along vortex lines. Long ago   Glaberson {\em et al.}\cite{glabL,glab} showed that this may course instability with respect to excitation of vortex array waves. Let us discuss  the threshold for the Glaberson--Johnson--Ostermeier  instability starting from the case of a uniform superfluid rotating with the angular velocity $\Omega$.

The general spectrum of the linear plane waves propagating in a rotating superfluid  $\propto e^{ipz+i\bm k \cdot \bm r -i\omega t}$  is \cite{RMP}
\be 
\omega^2 =(2\Omega +\nu_s p^2) \left(2\Omega\frac{p^2}{k^2+p^2}+\nu_s p^2+{c_T^2 k^2\over 2\Omega}\right),
  \el{spG}
where $p$ is the wavenumber along the $z$ axis, $\bm k$ is the wave vector in the $xy$ plane, and $c_T =\sqrt {\kappa \Omega/8\pi}$ is the Tkachenko-wave velocity determined by the shear rigidity of the vortex lattice. 
Further in this subsection we shall neglect the Tkachenko contribution since the calculation has shown that its effect is not essential.
The Landau critical  velocity for the superflow along the vortex line  for instability with respect to excitation of a wave with wave vector $(\bm k,p)$ is 
\be
v_c(k)={\omega\over p}=\sqrt{(2\Omega +\nu_s p^2) \left(\frac{2\Omega}{k^2+p^2}+\nu_s  \right)}.
     \ee    
The      minimum of this velocity at small $p \ll \sqrt{2\Omega /\nu_s}$ and large $k \gg \sqrt{2\Omega /\nu_s}$   is the critical velocity for the Glaberson--Johnson--Ostermeier instability:
\be
v_G=\sqrt{2\Omega\nu_s }.
     \el{OG}    
So the instability starts not from pure Kelvin waves ($p\gg k$) but from waves with $p\ll k$ propagating normally to the rotation axis. Still it would be interesting to discuss the instability for pure Kelvin waves $k=0$ when $v_c= (2\Omega +\nu_s p^2) /p$. Analyzing the stability of a single vortex in a rotating coordinate frame \citet{glab} obtained $v_c= (\Omega +\nu_s p^2) /p$. The factor 2 difference in the limit $p\to 0$ is due to long-range interaction between vortices. In contrast to the case of a single vortex, the HVBK theory of the vortex array  considers a uniform displacement of the {\em whole} vortex array. So any vortex feels the velocity induced not only  by  its own displacement but also by displacements of other vortices. This is valid until the wavenumber $p$ does not  exceed the inverse intervortex distance $1/r_v \sim \sqrt{\Omega /\kappa}$. If $p r_v \gg 1$ the velocity field induced by an oscillating vortex is cut off at the distance of the order the wavelength $2\pi/p$, and the gap of the Kelvin-wave spectrum becomes $\Omega$ as in the single-vortex case. The difference between the Kelvin-wave spectra for $p r_v \gg 1$ and $p r_v \ll 1$ was noticed by \citet{RG}  long ago.

Switching to instability in a twisted vortex bundle one should replace the uniform vorticity $2\Omega$ in the expression for $v_G$ by the space-dependent vorticity modulus $\omega_0(r)$, i.e., $v_G=\sqrt{\omega_0\nu_s }$. This approach is justified since $\omega_0(r)$ varies slowly at the scale $  \sqrt{\nu_s/2\Omega}$ at which instability sets on. The Glaberson--Johnson--Ostermeier critical velocity $v_G$ should be compared with the superfluid velocity $v_l=(v_\phi-\Omega r) s_\phi  +v_z s_z $ along the vortex lines in the rotating coordinate frame. For a large number vortices when $\Omega R^2 \gg \nu_s$ one can use calculations for continuous vorticity (Sec.~\ref{CV}). Then the twisted vortex bundle is stable only for rather weak twist satisfying the inequality
\be
Q^2 R^2 <{8 \nu_s \over \Omega R^2}.
      \el{GJO}

\section{Torsional oscillations of the vortex bundle}\label{TM}
\subsection{Phenomenological approach}

One should expect that weak stationary uniform twisting is related with some soft (Goldstone)  torsion mode related with axial symmetry of the vortex bundle. Let us start from the phenomenological approach to torsional oscillation: there are weak oscillations of the angular momentum density accompanied with the oscillating flux of the angular momentum. Restricting ourselves with a single degree of freedom related to a pair of canonically conjugated variables ``$z$ component  of the moment $m_z$ --the angle $\varphi$ of rotation around the axis $z$''  the phenomenological Hamiltonian is
\be
{\cal H} = \int \left({m_z^2\over 2I}+ {A \nabla_z \varphi^2\over 2}  \right)dz,  
  \ee   
where $I$ is the moment of inertia and $A$ is torsion stiffness. The Hamilton equations are
\bem
{\partial \varphi \over \partial t}= {\delta {\cal H}\over \delta m_z}= {m_z(z)\over I}, 
\nonumber \\
{\partial m_z \over \partial t}=- {\delta {\cal H}\over \delta \varphi}=\nabla_z\left( A \nabla_z \varphi \right).
  \eml{HE}
For a plane wave $\propto e^{ipz-i\omega t}$ propagating along the axis $z$ these equations yield the dispersion relation
\be 
\omega^2 =v_w^2 p^2={A\over I} p^2, 
     \el{sp}
where $v_w=\sqrt{A/I}$ is the velocity of the torsion mode.
\Eq{HE} confirms that the quantity $J_m=-A\nabla_z \varphi$ can be considered  as the angular momentum flux along the $z$ axis.

The effective Hamiltonian ${\cal H}$ can be derived from the Gibbs potential  of the liquid with the vortex bundle. Keeping in mind that  $Q=\nabla_z \varphi$ the torsion stiffness is determined by expansion  of $g$ with respect to $Q$. In the continuum model, Sec.  \ref{CV}, this gives $A=\pi \rho_s \Omega^2  R_0^6/3$ for the thin bundle $R_0\ll R$ and $A=\pi \rho_s \Omega^2  R^6/12$ for the thick bundle $R\approx R_0$. As for the inertial term it follows from the expansion of the Gibbs potential $g(\Omega')$ with respect to $\Omega'-\Omega =\partial \varphi /\partial t$. This yields the moment of inertia $I=\pi \rho_s R_0^4/2$ independently of the ratio $R/R_0$.
Then according to \eq{sp} the oscillation dispersion relation is
\be 
\omega^2 ={2\over  3}\Omega^2 R_0^2 p^2
     \el{frPh}
for $R_0 \ll R$ and
\be 
\omega^2 ={1\over  6}\Omega^2 R_0^2 p^2
     \el{frPh1}
for $R_0 \approx R$.

\subsection{Analysis based on the linear vortex dynamics} \label{Tk}

The simple analysis given above relied on the assumption that even in the excited state the bundle rotates as an ideal solid body with the velocity field given by \eq{clust}. The assumption is not self-evident, and moreover, not always valid as demonstrated below.    A more accurate approach must start from the vortex dynamics equations with proper boundary conditions. The linear hydrodynamic equations for a superfluid with vorticity were presented in Ref. \onlinecite{RMP}. For slow motion described by the plane wave $\propto e^{i\bm k \bm r +ipz -i\omega t}$    these equations are:
\bem
-i\omega v_{s\parallel}  -2\Omega v_{L\perp } -{i\omega k^2 \over p^2}v_{s\parallel} =0,  \nonumber \\
-i\omega v_{s\perp} +2\Omega v_{L\parallel } =0, 
             \eml{sys3}
\bem 
v_{L\parallel }=-i\omega u_\parallel =  v_{s\parallel}+{c_T^2k^2\over  2\Omega }u_\perp,
\nonumber \\
v_{L\perp}=-i\omega u_\perp \approx v_{s\perp},
   \eml{uph}
where $\bm u$ and $\bm v_L=-i\omega \bm u$ are the displacement and the velocity of the vortex lines  from their equilibrium positions in the $xy$ plane, and subscripts $\parallel$ and $\perp$ denote vector components in the $xy$ plane parallel and normal to the in-plane wave vector $\bm k$. The equations correspond to the dispersion relation given by \eq{spG} with the line-tension terms $\propto \nu_s$ neglected:
\be 
\omega^2 = 4\Omega^2\frac{p^2}{k^2+p^2}+c_T^2 k^2,
  \el{SIM}
The slow mode under consideration is a combination of the inertial wave (the first term $\propto \Omega^2$) and the Tkachenko wave (the second term $\propto c_T^2$). This combined mode (called the mixed mode in Ref. \onlinecite{RMP}) was crucial for interpretation of experiments on Tkachenko waves in $^4$He and on the slow mode  in $^3$He-B (see Refs.~\onlinecite{RMP} and  \onlinecite{ColModes}).    

The torsional oscillation mode is  a plane wave along the $z$ axis and an axisymmetric cylindric wave in the $xy$ plane. In the configurational space in the cylindric system of coordinate the equations of motion for the axisymmetric slow mode (no dependence on the azimuthal angle $\phi$) are 
\bem
{\partial ^2v_{sr}\over \partial z^2} -2\Omega {\partial ^2 u_{\phi }\over \partial z^2} +\left({\partial^2 v_{sr}\over \partial r^2}+{1\over r}{\partial v_{sr}\over \partial r}- { v_{sr} \over r^2}\right)=0,  \nonumber \\
{\partial u_\phi\over  \partial t} +2\Omega u_{r} =0,
\nonumber \\
{\partial u_r\over  \partial t}=v_{sr} -{c_T^2\over  2\Omega }\left({\partial^2 u_{\phi } \over \partial r^2}+{1\over r}{\partial u_{\phi }\over \partial r}- { u_{\phi } \over r^2}\right), 
   \eem 
where radial (subscript $r$) and azimuthal (subscript $\phi$) components correspond to
the longitudinal (subscript $\parallel$) and transverse (subscript $\perp$) components   in Eqs.~(\ref{sys3}) and Eqs.~(\ref{uph}) respectively. The general solution of these equation finite at the axis of the container is
\bem
u _\phi  = u_0 J_1(kr)e^{ipz-i\omega t}, ~~v_{sr} ={2\Omega p^2 u_0\over k^2+p^2}  J_1(kr)e^{ipz-i\omega t}, 
  \eml{solCon}
where $J_i(kr)$ are the Bessel functions of the first kind. The axial velocity 
\be
v_{sz} ={2\Omega k p\over k^2+p^2} i u_0 J_0(kr)e^{ipz-i\omega t}
       \el{vz}
is determined from the incompressibility condition 
\be
{\partial v_{sz} \over dz} +{\partial v_{sr} \over dr }+{v_{sr} \over r }=0.
       \ee 

The next step is to formulate proper boundary conditions. The first one is imposed on the liquid velocity and follows from the analysis of the vortex-free region, where the velocity is divergence- and curl-free. The general axisymmetric velocity field satisfying these conditions  is 
\bem 
v_{sr}(r) =[A  I_1(pr)-B K_1(pr)]e^{ip z},
~~
v_{s\phi}=0,~~v_{sz} (r)=i [A  I_0(pr)+B K_0(pr)]e^{ip z}.
   \eem 
Here   $I_0(pr)$ and $K_0(pr)$   are the modified Bessel functions of the first and the second kind.   The constant $A$ and $B$ are determined by the boundary condition at $r=R$:
\be 
v_{sr}(R) =[A  I_1(pR)-B K_1(pR)]e^{ip z}=0,
       \ee 
and by the continuity condition on the boundary of the vortex bundle $r=R_0$:
\bem 
 v_{sr}(R_0) =[A  I_1(pR_0)-B K_1(pR_0)]e^{ip z},
 \nonumber \\
 v_{sz}(R_0) =i [A  I_0(pR_0)+B K_0(pR_0)]e^{ip z} ,
       \eem
where $ v_{sr}(R_0)$ and $ v_{sz}(R_0)$ are the superfluid velocity components determined by Eqs.~(\ref{solCon}) and (\ref{vz}).

At $pR \ll 1$ this imposes the following condition on the velocity at the boundary of the vortex bundle:
\be 
iv_{sr}(R_0) -v_{sz}(R_0){p(R_0^2-R^2)\over 2 R_0}= 0.
    \el{BC}
For a thin bundle $R\gg R_0$ this reduces to the condition  $v_{sz}(R_0)=0$, while if the vortex bundle fills the whole container ($R- R_0 \ll R_0$), the radial velocity component must vanish: $v_{sr}(R_0)=0$. This reduces to the condition $2\pi \int_0^{R_0} v_{sz} r\,dr=0$ of zero axial mass flow used for the stationary twisted state.

The second boundary condition is imposed on the shear components of the vortex-lattice stress tensor, which must vanish at the boundary of the vortex bundle \cite{RMP}:
\bem
\sigma_{\phi r}(R_0)=\left.-\rho_s c_T^2\left[{\partial u_\phi \over \partial r} -{u_\phi\over r}\right] \right\vert_{r=R_0}=0.
   \eml{stress}

First let us look for a solution of the boundary problem neglecting the Tkachenko shear rigidity ($c_T \to 0$).  Then the boundary condition (\ref{stress}) is not relevant.   Let us focus on the case of a thin vortex bundle $R_0\ll R$.  Other cases can be treated similarly. If  $R_0\ll R$ the boundary condition $v_{sz}(R_0)=0$ requires that $J_0(kR_0)=0$. The  slow mode corresponds to the lowest nonzero root of the Bessel function: $k=2.405/R_0$. 
This yields the frequency of the slow torsion mode:
\be 
\omega^2 \approx  4\Omega^2\frac{p^2}{k^2}=0.692 \Omega^2 R_0^2 p^2.
  \ee 
Though this frequency only  slightly differs from the frequency in \eq{frPh} (numerical factor 0.692 vs. factor 0.667) obtained from the phenomenological approach the origin of the difference is worth of discussion. The solution of dynamical equations does not support the assumption of the phenomenological approach that the vortex bundle oscillates as an ideally solid body.   Indeed, \eq{solCon} shows that the azimuthal velocity $v_{s\phi}=-i\omega u_\phi$ is proportional to the Bessel function $J_1(kr)$ where $k\sim 1/R_0$ and is different from the solid-body rotation when $v_{s\phi} \propto r$. On the other hand, the question arises, whether ignoring of Tkachenko rigidity is justified in this case. If $p$ is small in the dispersion relation  (\ref{SIM}) the Tkachenko term $c_T^2k^2 \sim \kappa \Omega /R_0^2$ definitely exceeds the other term related with the classical inertial wave  and therefore cannot be neglected.

Thus we must take into account the Tkachenko rigidity and look for possible in-plane wavenumbers $k$ satisfying the dispersion relation (\ref{SIM}).  The latter is a bi-quadratic equation for $k$ so two values of $k^2$ are possible: 
\be
k_\pm^2 ={1\over 2}\left({\omega^2 \over c_T^2}-p^2\right) \pm \sqrt{ {1\over 4}\left({\omega^2 \over c_T^2}+p^2\right)^2-{4\Omega^2 \over c_T^2}p^2 }.
  \ee 
Since $\omega \sim \Omega R_0 p$  is proportional to $p$ [see Eqs.  (\ref{frPh}) and (\ref{frPh1})] but much larger than $c_T p$ this expression at  small $p$ reduces to 
\be
k_\pm^2={1\over 2}{\omega^2 \over c_T^2}\pm i {2\Omega  \over c_T}p.
   \el{spec}
It is important that $|k_\pm| $ is much larger than $p$ but still much smaller than the inverse radius $1/R_0$. The latter condition will allow to expand Bessel functions (see below).  The general solution of the boundary problem is a superposition of two modes with $k_+$ and $k_-$:
\bem
u_\phi= [u_+J_1(k_+r)+ u_-J_1(k_-r)],~~v_{sr} = 2\Omega p^2\left[u_+{J_1(k_+r) \over  k_+^2+p^2 }+u_-{J_1(k_-r) \over  k_-^2+p^2 }\right],
\nonumber \\
v_{Lr} ={\omega^2\over  \Omega} [u_+J_1(k_+r)+u_-J_1(k_-r)],~~
v_{sz} = i2\Omega p \left[u_+{k_+J_0(k_+r) \over  k_+^2+p^2 }+u_-{k_-J_0(k_-r) \over  k_-^2+p^2 }\right],
\nonumber \\
{du_\phi\over dr} -{u_\phi\over r} =- u_+ k_+J_2(k_+r)-u_- k_- J_2(k_-r),~~
\omega_\phi ={\partial v_r \over \partial z}-{\partial v_{sz} \over \partial r}= 2\Omega i pu_\phi.
  \eem 

The constants $u_+$ and $u_-$ are determined by the boundary conditions (\ref{BC}) and (\ref{stress}). In the case $R_0\ll R$ the determinant of these two linear equations vanishes at the condition
\bem
{(k_+^2+p^2)k_+ J_2(k_+R_0)\over  k_+J_0(k_+R_0)}-{(k_-^2+p^2)k_- J_2(k_-R_0)\over  k_-J_0(k_-R_0)}
\nonumber \\
\approx  k_+^4(1+k_+^2 R_0^2/ 6)-k_-^4(1+k_-^2 R_0^2/ 6) ={\omega^2 \over c_T^2}- {R_0^2\over 6}{4\Omega^2p^2 \over c_T^2}=0.
     \eml{TkSp}
Here the expression (\ref{spec})  for $k_\pm^2$ and the condition $k_\pm R_0 \ll 1$ were used. \Eq{TkSp} yields exactly the same dispersion relation (\ref{frPh}) as the phenomenological approach. It is interesting that this dispersion relation does not contain the circulation quantum, which determines the Tkachenko rigidity. Despite it the rigidity is important since it provides a solid body motion of the vortex bundle.

The analysis of the present section ignored mutual friction focusing on the $T=0$ limit. Meanwhile, whatever  weak mutual friction is, at very long wavelength of the torsion (slow) mode its spectrum transforms from sound-like to diffusive. For the clamped regime when the normal component co-rotates with the container as a single solid body (the case relevant for superfluid $^3$He) this was demonstrated in Sec.~VII.F of Ref.~\onlinecite{RMP}. The diffusive slow mode in $^3$He-B  was investigated experimentally and theoretically in Ref.~\onlinecite{ColModes}. In connection with the torsional oscillation of the vortex  bundle the diffusive slow mode was discussed by \citet{Twist}.

\section{The bundle terminating at the wall: propagation of the vortex front} \label{front}

\subsection{Single vortex line terminating at the lateral wall} \label{sinVor}

Before considering a vortex bundle terminating at the lateral wall (the main topic of this section, see Fig.~\ref{f1}) it is instructive to analyze a much simpler problem of a single vortex terminating at the lateral wall. This case allows an analytical solution, which helps to understand what may be going on in the vortex bundle. 

While without mutual friction the vortex line moves with the local velocity of the superfluid $\bm v_{sl}$ (see Sec.~\ref{QLT}), in the presence of mutual friction the  velocity of the vortex line depends on the mutual friction parameters $\alpha$ and $\alpha'$:
\be
\bm v_L=\bm v_{sl} + \alpha [\hat s \times (\bm v_n -\bm v_{sl})]-\alpha'[ \hat s \times  [ \hat s \times(\bm v_n -\bm v_{sl})]],
   \ee{} 
where $\hat s$ is the unit vector tangent to the vortex line. Studying the effect of mutual friction one must know the external force on the whole liquid (per unit length of a vortex line):
\be
\rho_s \kappa [\hat s \times (\bm v_L -\bm v_{sl})]=\rho_s \kappa\{-d[ \hat s \times  [ \hat s \times(\bm v_L -\bm v_n)]]+d' [\hat s \times (\bm v_L -\bm v_n)]\}.
   \el{forc} 
 Here the other friction parameters $d$ and $d'$ were introduced, which were used by \citet{Bevan}   and  are more convenient for the present analysis. They are connected with $\alpha$ and $\alpha'$ by the relations
  \be
 \alpha={d \over d^2+(1-d')^2},~~1-\alpha'={1-d' \over d^2+(1-d')^2}.
   \el{dd}

The vortex line goes along the container axis (the $z$ axis) deviating from it continuously starting from some height and eventually terminates at the lateral wall. The end segment of the vortex line connecting the axis and the lateral wall can be also considered as an elementary case of the vortex front. Stationary propagation of this vortex front along the $z$ axis supposes that the vortex moves as a solid body  with constant vertical velocity $v_f$, at the same time rotating around the  $z$ axis with angular velocity $\Omega_f$ different from the container angular velocity $\Omega$. The normal liquid corotates with the container. This means that  the relative velocity $\bm v_L -\bm v_n$ inside the front is fully defined and has only two components: the $z$ component $v_f$ and the azimuthal component $(\Omega_f-\Omega) r$.

For a single vortex in the local-induction approximation the superfluid velocity is determined by the line-tension force:
\be
[\hat s \times \bm v_{sl}] =-\nu_s \bm N.
     \el{curv}
Here  $ \bm N=d\hat s/dl$ is the curvature vector with its magnitude equal to the inverse curvature radius. Two functions $z(r)$ and $\phi(r)$ determine the shape of the vortex line in  cylindric coordinates $r,\phi,z$. Using Eq.~(\ref{curv}) and the expression for the curvature vector in the cylindric coordinate frame the vector equation   (\ref{forc}) leads to  two equations for the axial and the azimuthal components:
\be 
s_r \left\{ \Omega_f r + \nu_s{d\over dr}  {dz/dr \over [1+r^2(d\phi/dr)^2+(dz/dr)^2]^{1/2}}\right\} =(1-s_z^2)d v_f  +  (s_r d'-s_z s_\phi d) (\Omega_f -\Omega) r,
     \el{ax}
\be
s_r \left\{ - v_f +{\nu_s\over r}{d\over dr}  {r^2(d\phi/dr) \over [1+r^2(d\phi/dr)^2+(dz/dr)^2]^{1/2}} \right\} =(1-s_\phi^2)  d (\Omega_f -\Omega) r -(s_rd'+s_z  s_\phi d)v_f.
   \el{az}
 The third radial component of the force balance equation (\ref{forc}) is not an independent equation being a consequence of Eqs.  (\ref{ax}) and (\ref{az}).

Integrating Eqs.~ (\ref{ax}) and (\ref{az}) over the whole vortex line (keeping in mind that the line length element is $dl=dr / s_r= \sqrt{1+r^2(d\phi/dr)^2+(dz/dr)^2} dr$) and taking into account the boundary conditions  $dz/dr=\infty$ at $r=0$  and  $dz/dr=d\phi/dr=0$ at $r=R$ one obtains 
\be 
 \Omega_f {R^2 \over 2}- \nu_s=d_z  v_f R  +(d'-d_{z\phi} )(\Omega_f -\Omega) {R^2\over 2},
 \el{sh1}
 \be
- v_f{R^2\over 2}  =  d_\phi (\Omega_f -\Omega) {R^3\over 3} -(d'+d_{z\phi} )v_f{R^2\over 2},
   \el{sh2}
where three new mutual friction parameters related to the dissipative force $\propto d$ were introduced:
\be
d_z ={d\over R}\int_0^R {1- s_z^2 \over s_r}dr,~~d_\phi ={3d\over R^3}\int_0^R {1- s_\phi ^2 \over s_r}r^2\,dr,~~
d_{z\phi} ={2 d\over R^2}\int_0^R {s_zs_\phi \over s_r}r\,dr.
      \ee
\Eq{sh1} is the balance of axial forces   on the vortex, while \eq{sh2} is the balance of moments around the $z$ axis. This becomes evident if one rewrites them as 
\be
 e-\Omega_f  m_z = -\rho_s \kappa \left[d_z  v_f R  +(d'-d_{z\phi} )(\Omega_f -\Omega) {R^2\over 2}\right],
 \el{shh1}
 \be
v_f m_z  =-\rho_s \kappa \left[  d_\phi (\Omega_f -\Omega) {R^3\over 3} -(d'+d_{z\phi} )v_f{R^2\over 2} \right],
  \el{shh2}
where $e= \rho_s \kappa  \nu_s$ is the energy  and $m_z$ is the $z$ angular momentum per unit  length of straight vortex line far below the termination point respectively. For the vortex  coaxial with the container  and at the  distance $r_1$ from the container axis 
\be 
m_z=\rho_s \kappa {R^2-r_1^2\over 2}.
            \el{mom} 
For the axial vortex considered here $r_1=0$. The left-hand side of \eq{shh1} is a mechanical force on the vortex front, which is balanced by the axial friction force on the right-hand side. \Eq{shh2} demonstrates the balance between the moment  transferred to the liquid because of the vortex propagation (the left-hand side) and the mutual friction torque (the right-hand side). All terms in the balance equations except for the dissipative terms $\propto d$ do not depend on the shape of the vortex. In particular, the terms  $\propto v_f$ and $\Omega_f$ originate from the Magnus-force term $\rho_s \kappa [\hat s \times \bm v_L ] $ in \eq{forc}. Both the terms contain the angular momentum of the single vortex state per unit length.
   
The equations  of the linear and and the angular momenta balance lead to the balance of the total energy, which determines the energy dissipation:
\bem
{dE\over dt}= v_f  (e-\Omega_f  m_z) +(\Omega_f -\Omega) v_f m_z
\nonumber \\
=-\rho_s \kappa \left[d_z  v_f^2 R +d_\phi (\Omega_f -\Omega)^2 {R^3\over 3} -d_{z\phi} (\Omega_f -\Omega)v_f R^2\right] .
     \eml{rate}

Solving Eqs. (\ref{sh1}) and (\ref{sh2}) one obtains
\bem
 v_f =\frac{2d_\phi R(\Omega- \Omega_0)}{4d_z d_\phi+3[(1 -d') ^2-d_{z\phi}^2]}, 
  \nonumber \\
 \Omega_f -\Omega =-\frac{3(1-d'-d_{z\phi})(\Omega- \Omega_0)}{4d_z d_\phi+3[(1 -d') ^2-d_{z\phi}^2]} ,
       \eml{bal}
where $\Omega_0 = 2\nu_s /R^2$ is the critical angular velocity at which the Gibbs potentials of the single vortex state below the vortex front  and of the vortex-free state above the front are equal.
One cannot use these expressions directly   for determination of the propagation and the rotation velocities $v_f$ and $\Omega_f$ because  the mutual friction parameters $d_z$, $d_\phi$, and $d_{z\phi}$ depend on the vortex shape determined from the differential equations (\ref{ax}) and (\ref{az}).
However assuming that the friction force is weak, either because $d$ and $d'$ are small or because the difference $\Omega-\Omega_0$ is small one can determine the values $d_z$, $d_\phi$, and $d_{z\phi}$ using the shape of the vortex line in the state of equilibrium solid-body rotation together with the container and the normal liquid [Eq.~(13) in Ref.~\onlinecite{SN}]. The equilibrium assumes that $v_f=0$, $ \Omega=\Omega_f=\Omega_0$, and the vortex line lies in the axial plane:
\be
s_\phi=0,~~s_r= \frac{1} {\sqrt{1+(dz/dr)^2}}={r\sqrt{2R^2-r^2}\over R^2},~~s_z=\sqrt{1-s_r^2}={R^2-r^2\over R^2}.
   \ee
   Then 
\bem
d_z=d\int_0^1\rho\sqrt{2-\rho^2}d\rho={2\sqrt{2}-1\over 3}d=0.609 d ,
\nonumber \\
d_\phi =3d \int_0^1 {\rho\,d\rho \over\sqrt{2-\rho^2}}=3 (\sqrt{2}-1)d=1.24 d,~~d_{z\phi}=0.
   \eem 
Using these values in \eq{bal}
one obtains
\bem
 v_f  = \frac{2.48 d R(\Omega- \Omega_0)}{3.02 d^2+3(1 -d') ^2},
 \nonumber \\
 \Omega_f -\Omega =-\frac{3(1-d')(\Omega- \Omega_0)}{3.02 d^2+3(1 -d') ^2},
   \eml{balW}
the dissipation rate being
\bem
{dE\over dt}=
-\rho_s \kappa\frac{d_\phi(\Omega- \Omega_0) ^2 R^3}{4d_zd_\phi+3(1 -d') ^2}
=
-\rho_s \kappa\frac{1.24 d(\Omega- \Omega_0) ^2 R^3}{3.02 d^2+3(1 -d') ^2}.
             \eml{rateW}

The general expressions for the dissipation rate and the vortex front velocity $v_f$ [Eqs. (\ref{rate}) and (\ref{bal})] point out that these quantities are determined not by the  mutual friction parameter  $\alpha$ only, contrary to previous estimations in the literature \cite{EltProg}, and in general  the another parameter $\alpha'$ also affects the results.  However, in the limit of the weak friction force one may neglect a tiny difference between the factors 3 and 3.02 in the denominators of Eqs. (\ref{balW}) and (\ref{rateW}). Then according to \eq{dd} the dissipation rate and the vortex front depend only on $\alpha$. In particular, in this limit the vortex  front velocity is
\be
v_f\approx 0.83 \alpha R(\Omega- \Omega_0).
       \el{vf}
Recently the single-vortex front dynamics was numerically simulated on the basis of the Bio--Savart law \citep{KHT}. The results are in a qualitative agreement with those obtained here using the local-induction approximation. In particular,  expansion of the expression (11) of Ref. \onlinecite{KHT} for the front velocity (noted as $v_{Lz}$ there) with respect to $\Omega -\Omega_0$ gives $ v_f\approx 0.741 \alpha R(\Omega- \Omega_0)$, which does not differ essentially from  \eq{vf}.
A more quantitative comparison requires more numerical data for lower angular velocities $\Omega \sim \Omega_0$ while available data  \cite{KHT} focus on high angular velocities $\Omega \gg \Omega_0$.

Another useful approximation is to assume that the shape of  the end vortex segment in the front is close to a horizontal straight line between the axis ($r=0$) and the wall  ($r=R$). Then $s_z=s_\phi =0$ and $s_r=1$, which corresponds to $d_{z\phi}=0$ and $d_z=d_\phi =d$. In this case the final expression describing the vortex motion are 
\bem
 v_f  = \frac{2 d R(\Omega- \Omega_0)}{4 d^2+3(1 -d') ^2},
 \nonumber \\
 \Omega_f -\Omega =-\frac{3(1-d')(\Omega- \Omega_0)}{4 d^2+3(1 -d') ^2},
\nonumber \\
{dE\over dt}=
-\rho_s \kappa\frac{d(\Omega- \Omega_0) ^2 R^3}{4d^2+3(1 -d') ^2}.
             \eem
The difference with the exact solution for the weak-friction case is not so pronounced, and we shall exploit this approximation for a more complicated case of the vortex bundle.

\subsection{Propagation of the vortex front:  the energy and moment balance equations}

Let us consider now a bundle terminating at the lateral wall. The part of the bundle diverging to the wall is a vortex front separating the vortex-filled and the vortex-free parts of the container (Fig.~\ref{f1}). Motion of the front along the container axis is a transient process of vorticity penetration into a container in the spin-up experiments  \cite{EltProg}. Sometimes one can find stationary states of the bundle when the front does not move along the $z$ axis and the bundle and the  front (whorl) rotate as a solid body with the angular velocity determined from the thermodynamic analysis \cite{SN}. Here we  address the  case of a moving front.

 The front motion leads to change of the energy and the angular momentum and is accompanied
by mutual friction with the normal component moving rigidly with walls, or surface pinning and friction of vortex ends at rough surfaces of walls.  We look for the state with the vortex front moving 
with constant velocity $v_f$,  while the angular momentum necessary for this motion being supplied via the vortex pinning at the bottom and transferred to the front by the flux of the angular momentum $J_m$. The flux is related with the  twist $Q=\nabla_z \varphi$ of the vortex-bundle stem. Let us start from observation that a uniformly twisted bundle stem in the wake of the front moving with the velocity $v_f$ is possible  only if the front rotates with respect to the bundle stem with the relative angular velocity
\be
\Delta \Omega=\Omega_f-\Omega =v_f \nabla_z \varphi.
  \el{kin}
Only this purely kinematic relation provides that the moving front leaves behind it a uniformly twisted bundle in the absence  of vortex reconnections.

The rotation and the vertical motion of the front with respect to the solid-body rotating container with the normal fluid  lead to friction.  The balance of the angular momentum around the axis $z$ during the front motion is
\be 
 m_z v_f +J_m =T_{fr}.
          \el{BalM}
The equation tells that the angular momentum brought by the angular-momentum flux $\left.  \left. J_m=-\partial e/\partial \nabla_z \varphi \right|_{m_z}=-\partial g/\partial \nabla_z \varphi \right |_{\Omega}$ is compensated by the growth of the total angular momentum (the term $m_z v_f$) due to front propagation and by the friction torque $T_{fr}$. 

The balance of the linear momentum along the axis $z$, as we shall demonstrate below, is
\be 
e - \Omega_f  m_z - \nabla_z \varphi J_m=F_{fr}.
  \el{force}
The left-hand side of the equation is a driving {\em mechanical} force on the front balanced by the friction force $F_{fr}$.  
Note that elongation of the bundle due to front propagation does not lead to variation of the linear momentum, because the latter is exactly zero above and below the front. So the mechanical force does not contain a term proportional to $v_f$ similar to the term  $ m_z v_f$ in the balance equation (\ref{BalM}) for the angular momentum.

It is important also to consider the balance of the energy:
\be
(e - \Omega m_z)v_f= -(\Omega_f-\Omega)T_{fr}-v_f F_{fr}.
   \el{BalE}
The left-hand side is the rate of the energy variation due to front propagation whereas the right-hand side is the dissipation rate. The Gibbs potential density $g=e - \Omega m_z$, which appears in the left-hand side of \eq{BalE}, has also the dimensionality of a force and may be called {\em effective} force. It  differs from the mechanical force in the left-hand side of \eq{force} because the vortex front not only moves along the container axis, but also rotates. For an untwisted bundle ($\nabla_z \varphi=0$) studied in Ref. \onlinecite{SN}  the two forces coincide. The  balance equations (\ref{BalM}) and  (\ref{force}) for the  angular and  linear   momenta of the vortex bundle differ from the similar balance equations (\ref{shh1}) and (\ref{shh2}) for the single vortex by the presence of terms containing the angular momentum flux $J_m$.

The balance equations (\ref{BalM}) and (\ref{BalE}) for the angular momentum and the energy seem evident. On the other hand, \eq{force} of the linear momentum balance needs a justification. The simplest argument in its favor  follows from the fact the three balance equations (\ref{BalM})--(\ref{BalE}) are not independent, and it is easy to check  that \eq{force} follows from Eqs.~(\ref{BalM}) and (\ref{BalE}). If one believes the two, one must accept the third one. It is possible also to derive 
Eqs.~(\ref{force}) and (\ref{BalM}) directly from the equations of hydrodynamics.   

All terms on the leftt-hand side of the equations, i.e, those not connected with mutual friction, do not depend on the velocity and vorticity distribution inside the vortex front, but only on the distribution inside the bundle stem far below the front. Let us demonstrate this for terms originating from the Magnus-force term $\rho_s \kappa [\hat s \times \bm v_L ] $ in \eq{forc}. In Sec.~\ref{sinVor} it was shown that the contributions of this term to the linear and the angular momentum balance of a single vortex are $\Omega_f m_z$ and $-v_f m_z$ respectively, where $m_z$ is the $z$ angular momentum per unit length given by \eq{mom}. For the vortex bundle one should find the summary effect of all vortices in the bundle. This requires integration of the vortex distribution over the bundle cross-section  far below the vortex front. In particular,  the Magnus force contribution to the linear momentum balance equation is
\be
2\pi \kappa \rho_s \Omega_f \int_0^R {R^2-r_1^2\over 2}  n_v(r_1) r_1\,dr_1 = 2\pi  \Omega_f \int_0^R r^2 v_{s\phi} (r)dr =\Omega_f m_z.
\nonumber 
   \ee    
Here $m_z$ is the angular momentum per unit length of the vortex bundle far below the vortex front,  and 
the relation 
\be
\kappa n_v(r) = {1\over r }{d(r  v_{s\phi})\over dr}
  \ee      
between the vortex density $n_v$ and the azimuthal superfluid velocity $v_{s\phi}$ was used. Similarly one can check that the Magnus force contribution to the angular momentum balance equation is $-v_f m_z$.

However, in order to derive the whole balance equations it is more convenient to start from the original hydrodynamic Euler equation for the superfluid component. For the sake of simplicity we restrict the derivation with the continuous-vorticity limit since its generalization on the case with line-tension effect and the mutual friction  force is straightforward. Then the Euler equation is
\be
{\partial \bm v_s \over\partial t} +(v_{sj} \nabla _j) \bm v_s+\bm \nabla \mu=0.
       \ee{} 
The $r$ and $z$ components of this vector equation in the cylindric coordinates $(r,\phi,z)$ for the axisymmetric velocity field (no  dependence on $\phi$) are 
\be
{\partial v_{sr}\over\partial t} + v _{sr} {\partial v_{sr}\over \partial r} + v _z {\partial v_{sr}\over \partial z}-{v_{s\phi}^2\over r}+{\partial \mu\over \partial r}=0,
    \el{rs}
    \be
{\partial v_{sz} \over\partial t} + v _{sr}{\partial v_{sz}\over \partial r} + v _z {\partial v_{sz} \over \partial z}+{\partial \mu \over \partial z}
=0.
     \el{zs}
For a stationary moving vortex front $\partial v_{sr}/\partial t=-v_f \partial v_{sr}/\partial z$ and $\partial v_{sz}/\partial t=-v_f \partial v_{sz}/\partial z$.
Choosing  the chemical potential $\mu$ in the vortex-free region  far above the vortex front as a reference point, i.e., assuming that $\mu=0$ there, one can integrate \eq{zs} for  the $z$ component along the container axis $r=0$. This yields the chemical potential at the container  axis far below the vortex front ($z=-\infty$):
\be
\mu(0,-\infty) = \int _{-\infty }^\infty dz\left(v_f {\partial v_{sz} \over \partial z} -{1\over 2}  {\partial v_{sz}^2 \over \partial z}\right)
=-v_f v_{sz}(0,-\infty)  -{ v_{sz}(0,-\infty)^2\over 2}.
   \ee
As a next step one can calculate dependence of $\mu(r)$ on the distance $r$ from the axis below the vortex front (further the argument $z=-\infty$ is omitted) integrating \eq{rs}:
\be
\mu(r)= v_f v_{sz}(0) -{v_{sz}^2(0)\over 2}+\int_0^r v_{s\phi}(r_1)^2{dr_1\over r_1}. 
   \ee
Now one can determine the total variation of the linear momentum of the liquid integrating the momentum flux component $\Pi_{zz}=P+ v_{sz}^2$ [\eq{flux}] over the cross-section of the bundle stem bearing in mind that this is a momentum only of the superfluid component and the pressure is $P=\rho_s \mu$:
\bem
2\pi \int_0^R \Pi_{zz} r\,dr=\pi R^2\left[ v_f v_{sz}(0) -{v_{sz}^2(0)\over 2}\right] + \pi \int_0^Rv_{sz}^2r\,dr +2\pi \int_0^R\left[\int_0^r  v_{s\phi}^2{dr_1 \over r_1}\right]r\,dr
\nonumber \\
=\pi R^2\left[ v_f v_{sz}(0) -{v_{sz}^2(0)\over 2}\right] +2 \pi \int_0^Rv_{sz}^2r\,dr +\pi R^2 \int_0^R v_{s\phi}^2{dr \over r}-\pi \int_0^Rv_{s\phi}^2r\,dr.
   \eml{eul}
Calculating integrals for the velocity field in the twisted bundle in the continuous-vorticity limit [Eqs.~(\ref{Thun}) and (\ref{v0})] and using the kinematic relation (\ref{kin}) one can check that \eq{eul} yields exactly the left-hand side of the linear momentum balance equation (\ref{force}).

Next let us calculate the total friction force and   the total friction torque integrating the mutual friction force  per unit volume [cf. \eq{forc}],
\be
\bm f_{fr}= d\omega_0 \left\{ (\bm v_L -\bm v_n) -\hat s [\hat s \cdot (\bm v_L -\bm v_n)]\right\} +d' \omega_0 [\hat s \times (\bm v_L -\bm v_n)].
\el{vfz}
over the whole vortex liquid:
\bem
F_{fr}=2\pi  \rho_s \kappa \int_{-\infty}^\infty dz\int _0^R\omega_0\left\{ d\left[  v_f      (1 -s_z  ^2) -s_z s_\phi(\Omega_f-\Omega)r\right]+d' s_r (\Omega_f-\Omega)r \right\}r\,dr
\nonumber \\
=2\pi  \rho_s \kappa \int_{-\infty}^\infty dz\int _0^R\left\{\omega_0 d\left[  v_f      (1 -s_z  ^2) -s_z s_\phi(\Omega_f-\Omega)r\right]-d' {\partial v_{s\phi}\over \partial z} (\Omega_f-\Omega)r \right\}r\,dr
\nonumber \\
=d v_f m_F + d'(\Omega_f-\Omega) m_z,
\nonumber \\
T_{fr}=2\pi  \rho_s \kappa \int_{-\infty}^\infty dz\int _0^R\omega_0\left\{ d\left[  (\Omega_f-\Omega)r     (1 -s_\phi  ^2) -s_z s_\phi v_f \right]-d' s_r v_f \right\}r^2\,dr
\nonumber \\
=d(\Omega_f-\Omega)m_T-d' v_fm_z.
   \eem    
Here $\omega_0$ is the absolute value of the vorticity vector $\bm \omega_0=\bm \nabla \times \bm v_s$ and 
\bem
m_F=2\pi   \rho_s \kappa \int_{-\infty}^\infty dz\int _0^R\omega_0   (1 -s_z  ^2 -s_z s_\phi Q r )r\,dr,
\nonumber \\
m_T=2\pi  \rho_s \kappa \int_{-\infty}^\infty dz\int _0^R\omega_0 \left( 1 -s_\phi  ^2 -{s_z s_\phi \over Qr}  \right)r^3\,dr
    \eml{mm}
are the effective  moments related with the dissipative parameter $d$. Only the vortex front region contributes to the bulk integrals, since  there is no vorticity in the vortex-free region above the front, while below the front the relative velocity $\bm v_L-\bm v_n$ has no component normal to the vortex lines and the integrand also vanishes.

The dissipation rate depends only on the dissipative mutual friction parameter $d$:
\be
(\Omega_f-\Omega)T_{fr}+v_f F_{fr}=d[ v_f^2m_F+(\Omega_f-\Omega)^2m_T]=d  v_f^2(m_F+\nabla_z \varphi^2m_T).
   \ee 

The explicit expressions for the friction force and the friction torque allow to derive from the balance equations the relations determining the front velocity $v_f$ and the twist $Q=\nabla_z \varphi$:
\be
{1-d'\over d}={1-\alpha '\over \alpha}=-{ m_TQ\over m_z}+J_m{m_F+Q^2 m_T\over (m_z\Omega-e)m_z},~~v_f=\frac{    m_z\Omega -e}{d(m_F+m_TQ^2)}.
   \el{Fin}

As in the single-vortex case, the dissipative forces depend on the velocity field inside the front, and in order to find the moments $m_F$ and $m_T$ we approximate the vortex line in the front by straight line segments normal to the axis. This means that $s_r=1$ and $s_z=s_\phi=0$, and $\omega_0 = -\partial v_\phi /\partial z$. After integration over $z$ \eq{mm} reduces to
\be
m_F = 2\pi \int_0^R  v_{s\phi} (r)r\,dr, ~~ m_T = 2\pi \int_0^R  v_{s\phi} (r)r^3\,dr.
  \ee

The further analysis restricts itself with the case of the bundle occupying the whole cross-section of the container ($R\approx R_0$).  Let us neglect first the line-tension effect using the relations given in Sec.~\ref{CV}. 
In the limit of strong   dissipative friction parameter $d$   compared to $1-d'$
 $v_f$ and $Q$ are expected to be small and the expansion in $Q$ reduces  \eq{Fin} to
\bem
{1-\alpha '\over \alpha} \rightarrow {56\over 45}QR ,~~v_f\rightarrow   \frac{3\Omega R} { 8d}. 
        \eml{HT}

One may compare this result with experimental measurements of the twist $Q$ and the vortex front velocity $v_f$ at temperatures higher than about 0.5 $T_c$. Figures 15 and 17 in Ref.~\onlinecite{EltProg} show that $Q$ grows and $v_f$ drops with decreasing temperature. This qualitatively agrees with  \eq{HT}
since both $d$ and $(1-\alpha')/\alpha = (1-d')/d$ grow with decreasing temperature according to measurements of  \citet{Bevan} for $^3$He-B at $T>0.6$ $T_c$ (see their Figs. 5, 6, 8, and 9).  The results of the experiment \cite{EltProg} and the numerical calculations of the HVBH equations \cite{EltPr} at high temperatures  were described by the expression $v_f \approx \alpha \Omega R$.  It is worthwhile of noting that at $\alpha \gg 1-\alpha'$ there is no difference between $\alpha $ and $1/d$, so the quoted result differs from  \eq{HT} only by the factor 3/8. It maybe explained by inaccuracy of our assumption on the shape of the bundle inside the vortex front.

From the position of the nowadays discussions of the $T=0$ limit  it is interesting to consider the opposite case of vanishing mutual friction $d,d'\to 0$. In this limit both the front velocity $v_f$ and $Q$ grow and 
\bem
{1-\alpha '\over \alpha} \rightarrow {2\over 3}QR ,~~v_f\rightarrow   \frac{3\Omega } { 4 dQ^2 R} =
 \frac{d\Omega R} {3}.
   \eml{WF}
Thus with vanishing mutual friction the twist  become extremely large, while  the vortex front velocity $v_f$ vanishes in this limit. At the same time the relative angular velocity $\Omega_f -\Omega =Q v_f$ remains finite and the front rotation is the most important source of energy dissipation. However, one should remember that friction will not fully disappear even in the $T=0$ limit: surface friction of vortex ends moving along a rough wall would restrict the velocity of the vortex front, as mutual friction does at $T>0$. Even a more serious problem with this limit is the Glaberson--Johnson--Ostermeier instability of the laminar regime at strong twist  demonstrated in Sec. \ref{GlabIn}.

According to Sec.~\ref{QLT} the joint effect of line tension and twist drives the system close  to the state where $g=e-\Omega  m_z$ vanishes. If the system reaches this state \eq{Fin} has a solution without mutual friction with the twist $Q$ determined from the condition $g=e-\Omega m_z=0$ and with the front velocity $v_f = J_m/m_z$. For large numbers of vortices the state is possible only for large $QR$, and using the expressions of Sec.~\ref{QLT} in this limit \eq{Fin} yields:
\be
Q= {\Omega R\over 2\nu_s}, ~~v_f={3\Omega \over 4Q \ln(QR)}.
   \ee 
The Glaberson--Johnson--Ostermeier instability at strong twist also puts  possibility to reach this state in the laminar regime in question. However,  this option can revive above the instability threshold, where strong oscillations can increase the energy $e$ allowing to reach the condition $e-\Omega  m_z=0$ at larger vortex number and weaker twists (see  discussion in the next section).

\section{Discussion and conclusions} \label{concl}

This work analyzed dynamics of twisted vortex bundles in rotating superfluids revealed in recent spin-up experiments on superfluid $^3$He-B  \cite{Elt07,EltProg,Elt11}. As the first step the linear dynamics was investigated, which demonstrated that the torsion oscillation mode involving weak bundle twisting is a particular case of the linear vortex dynamics of the  slow vortex mode. This mode was investigated in the past in connection with observation of the Tkachenko wave in superfluid $^4$He (Ref. \onlinecite{RMP}) and the experiments on the slow vortex relaxation in superfluid $^3$He-B (Ref.~\onlinecite{ColModes}). The strongly twisted bundle was also investigated, but it was demonstrated that the Glaberson--Johnson--Ostermeier instability prevents  reaching the strong-twist case in the laminar regime.  

The analysis addressed also a twisted vortex bundle terminating at a container lateral wall. The segment of the bundle diverging to the wall (vortex front, or whorl) is propagating along and is rotating around  the container axis.  The analysis starts from the case of a single vortex terminating at the wall,  which allows an analytic solution for a weak force driving the vortex along the container axis. The three equations for the balance of the linear and angular momenta and for the energy were derived, which were further generalized on the case of the vortex bundle. It was supposed that the vortex front propagation occurs in the laminar regime, without vortex reconnections. This provides a kinematic relation between the linear velocity and the angular velocity of the front and   allows to find the parameters of the vortex-front motion. In this equations only the dissipative components of the mutual friction force and torque require knowledge of the velocity and the vorticity distribution inside the vortex front, and their contributions to the balance equations were estimated using simple assumptions on the vortex line shape inside the front.

The analysis of the vortex-front propagation was performed under the assumption that the vortex bundle below the front fills the whole container cross-section. But vortex-line tension always leads to the vortex-free region near the container lateral wall \cite{Khalatnikov2000,RMP} at the equilibrium. So the analysis was done for the number of vortices exceeding the equilibrium vortex number. In the experiments on the vortex-front propagation one may expect that the  number of vortices in the bundle is determined not by the condition of equilibrium but the conditions of creation of the vortex bundle at the container bottom, and this number could be more or less than the equilibrium value. Knowing this number one can generalize the present analysis including the effect of the vortex-free region. This effect was neglected  for the sake of simplicity, but taking it into account would result in more accurate numerical coefficients in our final expressions.

Another essential assumption of the analysis was strong pinning of the vortex bundle at the  bottom, which provided the constant angular-momentum flux towards the vortex front. Though pinning at the lower end of the bundle is quite possible, especially if the ``bottom'' is in fact an interface separating the B phase from the A phase with the periodical vortex structure, one may address also the case without pinning. Then coupling between the vortex bundle and the container can be provided only by mutual friction. This would slightly complicate the analysis since the  angular-momentum flux proportional to the twist  will be not uniform along the container axis. The numerical calculations by \citet{EltPr}, who neglected pinning at the bottom, showed that the twist varied behind the vortex front rather slowly, nevertheless. Then one may use the present analysis assuming that the bundle twist in the theory is the twist well behind the vortex front but still rather far from the bottom.

The important question, which attracted   great attention in the literature on superfluid vortex dynamics nowadays, is: What is dynamics of the vortex front  in the limit of vanishing mutual friction (the $T=0$ limit)? Physical and numerical experiments provided evidence that  at low temperatures the laminar regime transforms to the turbulent regime. The Glaberson--Johnson--Ostermeier instability of the laminar regime at growing twist can be considered as a precursor of  this transformation. The instability condition is determined by the twist $Q$ whereas earlier they supposed \cite{Finne,VolRe,Elt10} that the transition to turbulence is governed by the ratio $1-\alpha'/\alpha$, which is the quality factor for Kelvin waves. However, the twist  $Q$ is connected with $1-\alpha'/\alpha$, and using this connection [\eq{HT}] the laminar-regime stability condition  given by the inequality (\ref{GJO}) can be rewritten as 
\be
{1-\alpha'\over \alpha} <3.5\sqrt{ \nu_s \over \Omega R^2}.
      \ee
The right-hand side of the inequality is rather small at a large number of vortices. So the Glaberson--Johnson--Ostermeier threshold predicts small values  of $1-\alpha'/\alpha$ at the transition, while in the experiments evidences for turbulence appear at $1-\alpha'/\alpha$ of order unity or more \cite{Finne,Elt10}. Probably this means that there is an intermediate stage between the laminar and turbulent regime characterized by large vortex array fluctuations but still without essential number of reconnections. Previously the effect of reconnections was investigated numerically. According to \citet{Elt11}, reconnections, which were registered in numerical experiments, become essential only below 0.3$T_c$. Meanwhile, the laminar regime apparently becomes unstable at higher temperatures.

The analysis of the intermediate regime is beyond the scope of the present work restricted with the laminar regime. But one may speculate what results of the present analysis obtained for the laminar regime could be retained in the new regime. Since reconnections are still not essential the kinematic relation (\ref{kin}) remains in force. This validates our treatment of the balance equations, but with energy, forces and torques recalculated taking into account large fluctuations  triggered by the Glaberson--Johnson--Ostermeier instability.
It is natural to expect that fluctuations increase the energy of the bundle behind the front. This would facilitate  reaching the condition $e-\Omega m_z=0$, which allows the vortex front motion  without friction at $T\to 0$  at vortex numbers larger than those in the laminar regime.  If it were realized it would explain observed saturation of the vortex-front velocity dependence at low temperatures.

\section*{ACKNOWLEDGMENTS}

The author appreciates interesting discussions with Vladimir Eltsov, Bill Glaberson,  Janne Karim\"aki, Matti Krusius,  Sergey Nemirovskii, and Erkki Thuneberg. Special thanks to  Risto H\"anninen, who found some missing factors in the previous version of this work.
The work was supported by the 7th European Community  Framework Program  under grant agreement  MICROKELVIN and by
the grant of the Israel Academy of Sciences and Humanities.





%


\begin{thebibliography}{20}%
\makeatletter
\providecommand \@ifxundefined [1]{%
 \@ifx{#1\undefined}
}%
\providecommand \@ifnum [1]{%
 \ifnum #1\expandafter \@firstoftwo
 \else \expandafter \@secondoftwo
 \fi
}%
\providecommand \@ifx [1]{%
 \ifx #1\expandafter \@firstoftwo
 \else \expandafter \@secondoftwo
 \fi
}%
\providecommand \natexlab [1]{#1}%
\providecommand \enquote  [1]{``#1''}%
\providecommand \bibnamefont  [1]{#1}%
\providecommand \bibfnamefont [1]{#1}%
\providecommand \citenamefont [1]{#1}%
\providecommand \href@noop [0]{\@secondoftwo}%
\providecommand \href [0]{\begingroup \@sanitize@url \@href}%
\providecommand \@href[1]{\@@startlink{#1}\@@href}%
\providecommand \@@href[1]{\endgroup#1\@@endlink}%
\providecommand \@sanitize@url [0]{\catcode `\\12\catcode `\$12\catcode
  `\&12\catcode `\#12\catcode `\^12\catcode `\_12\catcode `\%12\relax}%
\providecommand \@@startlink[1]{}%
\providecommand \@@endlink[0]{}%
\providecommand \url  [0]{\begingroup\@sanitize@url \@url }%
\providecommand \@url [1]{\endgroup\@href {#1}{\urlprefix }}%
\providecommand \urlprefix  [0]{URL }%
\providecommand \Eprint [0]{\href }%
\providecommand \doibase [0]{http://dx.doi.org/}%
\providecommand \selectlanguage [0]{\@gobble}%
\providecommand \bibinfo  [0]{\@secondoftwo}%
\providecommand \bibfield  [0]{\@secondoftwo}%
\providecommand \translation [1]{[#1]}%
\providecommand \BibitemOpen [0]{}%
\providecommand \bibitemStop [0]{}%
\providecommand \bibitemNoStop [0]{.\EOS\space}%
\providecommand \EOS [0]{\spacefactor3000\relax}%
\providecommand \BibitemShut  [1]{\csname bibitem#1\endcsname}%
\let\auto@bib@innerbib\@empty
\bibitem [{\citenamefont {Saffman}(1995)}]{Saf}%
  \BibitemOpen
  \bibfield  {author} {\bibinfo {author} {\bibfnamefont {P.~G.}\ \bibnamefont
  {Saffman}},\ }\href@noop {} {\emph {\bibinfo {title} {Vortex dynamics}}}\
  (\bibinfo  {publisher} {Cambridge University Press},\ \bibinfo {year}
  {1995})\BibitemShut {NoStop}%
\bibitem [{\citenamefont {Donnelly}(1991)}]{D}%
  \BibitemOpen
  \bibfield  {author} {\bibinfo {author} {\bibfnamefont {R.~J.}\ \bibnamefont
  {Donnelly}},\ }\href@noop {} {\emph {\bibinfo {title} {Quantized vortices in
  helium II}}}\ (\bibinfo  {publisher} {Cambridge University Press},\ \bibinfo
  {year} {1991})\BibitemShut {NoStop}%
\bibitem [{\citenamefont {Alekseenko}\ \emph {et~al.}(2007)\citenamefont
  {Alekseenko}, \citenamefont {Kuibin},\ and\ \citenamefont {Okulov}}]{Oku}%
  \BibitemOpen
  \bibfield  {author} {\bibinfo {author} {\bibfnamefont {S.~V.}\ \bibnamefont
  {Alekseenko}}, \bibinfo {author} {\bibfnamefont {P.~A.}\ \bibnamefont
  {Kuibin}}, \ and\ \bibinfo {author} {\bibfnamefont {V.~L.}\ \bibnamefont
  {Okulov}},\ }\href@noop {} {\emph {\bibinfo {title} {Theory of Concentrated
  Vortices}}}\ (\bibinfo  {publisher} {Springer},\ \bibinfo {year}
  {2007})\BibitemShut {NoStop}%
\bibitem [{\citenamefont {Eltsov}\ \emph {et~al.}(2006)\citenamefont {Eltsov},
  \citenamefont {Finne}, \citenamefont {H\"anninen}, \citenamefont {Kopu},
  \citenamefont {Krusius}, \citenamefont {Tsubota},\ and\ \citenamefont
  {Thuneberg}}]{Twist}%
  \BibitemOpen
  \bibfield  {author} {\bibinfo {author} {\bibfnamefont {V.~B.}\ \bibnamefont
  {Eltsov}}, \bibinfo {author} {\bibfnamefont {A.~P.}\ \bibnamefont {Finne}},
  \bibinfo {author} {\bibfnamefont {R.}~\bibnamefont {H\"anninen}}, \bibinfo
  {author} {\bibfnamefont {J.}~\bibnamefont {Kopu}}, \bibinfo {author}
  {\bibfnamefont {M.}~\bibnamefont {Krusius}}, \bibinfo {author} {\bibfnamefont
  {M.}~\bibnamefont {Tsubota}}, \ and\ \bibinfo {author} {\bibfnamefont
  {E.~V.}\ \bibnamefont {Thuneberg}},\ }\href@noop {} {\bibfield  {journal}
  {\bibinfo  {journal} {Phys. Rev. Lett.}\ }\textbf {\bibinfo {volume} {96}},\
  \bibinfo {pages} {215302} (\bibinfo {year} {2006})}\BibitemShut {NoStop}%
\bibitem [{\citenamefont {Finne}\ \emph {et~al.}(2003)\citenamefont {Finne},
  \citenamefont {Araki}, \citenamefont {Blaauwgeers}, \citenamefont {Eltsov},
  \citenamefont {Kopnin}, \citenamefont {M}, \citenamefont {Skrbek},
  \citenamefont {Tsubota},\ and\ \citenamefont {Volovik}}]{Finne}%
  \BibitemOpen
  \bibfield  {author} {\bibinfo {author} {\bibfnamefont {A.~P.}\ \bibnamefont
  {Finne}}, \bibinfo {author} {\bibfnamefont {T.}~\bibnamefont {Araki}},
  \bibinfo {author} {\bibfnamefont {R.}~\bibnamefont {Blaauwgeers}}, \bibinfo
  {author} {\bibfnamefont {V.~B.}\ \bibnamefont {Eltsov}}, \bibinfo {author}
  {\bibfnamefont {N.~B.}\ \bibnamefont {Kopnin}}, \bibinfo {author}
  {\bibnamefont {M}}, \bibinfo {author} {\bibfnamefont {L.}~\bibnamefont
  {Skrbek}}, \bibinfo {author} {\bibfnamefont {M.}~\bibnamefont {Tsubota}}, \
  and\ \bibinfo {author} {\bibfnamefont {G.~E.}\ \bibnamefont {Volovik}},\
  }\href@noop {} {\bibfield  {journal} {\bibinfo  {journal} {Nature}\ }\textbf
  {\bibinfo {volume} {424}},\ \bibinfo {pages} {1022} (\bibinfo {year}
  {2003})}\BibitemShut {NoStop}%
\bibitem [{\citenamefont {Volovik}(2003)}]{VolRe}%
  \BibitemOpen
  \bibfield  {author} {\bibinfo {author} {\bibfnamefont {G.~E.}\ \bibnamefont
  {Volovik}},\ }\href@noop {} {\bibfield  {journal} {\bibinfo  {journal}
  {Pis'ma v ZhETF}\ }\textbf {\bibinfo {volume} {78}},\ \bibinfo {pages} {1021}
  (\bibinfo {year} {2003})},\ \bibinfo {note} {[JETP Letters, {\bf 78}, 533
  (2003)]}\BibitemShut {NoStop}%
\bibitem [{\citenamefont {Eltsov}\ \emph {et~al.}(2007)\citenamefont {Eltsov},
  \citenamefont {Golov}, \citenamefont {de~Graaf}, \citenamefont {H\"anninen},
  \citenamefont {Krusius}, \citenamefont {L'vov},\ and\ \citenamefont
  {Solntsev}}]{Elt07}%
  \BibitemOpen
  \bibfield  {author} {\bibinfo {author} {\bibfnamefont {V.~B.}\ \bibnamefont
  {Eltsov}}, \bibinfo {author} {\bibfnamefont {A.~I.}\ \bibnamefont {Golov}},
  \bibinfo {author} {\bibfnamefont {R.}~\bibnamefont {de~Graaf}}, \bibinfo
  {author} {\bibfnamefont {R.}~\bibnamefont {H\"anninen}}, \bibinfo {author}
  {\bibfnamefont {M.}~\bibnamefont {Krusius}}, \bibinfo {author} {\bibfnamefont
  {V.~S.}\ \bibnamefont {L'vov}}, \ and\ \bibinfo {author} {\bibfnamefont
  {R.~E.}\ \bibnamefont {Solntsev}},\ }\href@noop {} {\bibfield  {journal}
  {\bibinfo  {journal} {Phys. Rev. Lett.}\ }\textbf {\bibinfo {volume} {99}},\
  \bibinfo {pages} {265301} (\bibinfo {year} {2007})}\BibitemShut {NoStop}%
\bibitem [{\citenamefont {Eltsov}\ \emph {et~al.}(2010)\citenamefont {Eltsov},
  \citenamefont {de~Graaf}, \citenamefont {Heikkinen}, \citenamefont {Hosio},
  \citenamefont {H\"anninen}, \citenamefont {Krusius},\ and\ \citenamefont
  {L'vov}}]{Elt10}%
  \BibitemOpen
  \bibfield  {author} {\bibinfo {author} {\bibfnamefont {V.~B.}\ \bibnamefont
  {Eltsov}}, \bibinfo {author} {\bibfnamefont {R.}~\bibnamefont {de~Graaf}},
  \bibinfo {author} {\bibfnamefont {P.~J.}\ \bibnamefont {Heikkinen}}, \bibinfo
  {author} {\bibfnamefont {J.~J.}\ \bibnamefont {Hosio}}, \bibinfo {author}
  {\bibfnamefont {R.}~\bibnamefont {H\"anninen}}, \bibinfo {author}
  {\bibfnamefont {M.}~\bibnamefont {Krusius}}, \ and\ \bibinfo {author}
  {\bibfnamefont {V.~S.}\ \bibnamefont {L'vov}},\ }\href@noop {} {\bibfield
  {journal} {\bibinfo  {journal} {Phys. Rev. Lett.}\ }\textbf {\bibinfo
  {volume} {105}},\ \bibinfo {pages} {125301} (\bibinfo {year}
  {2010})}\BibitemShut {NoStop}%
\bibitem [{\citenamefont {Eltsov}\ \emph {et~al.}(2009)\citenamefont {Eltsov},
  \citenamefont {de~Graaf}, \citenamefont {H{\"a}nninen}, \citenamefont
  {Krusius}, \citenamefont {Solntsev}, \citenamefont {L'vov}, \citenamefont
  {Golov},\ and\ \citenamefont {Walmsley}}]{EltProg}%
  \BibitemOpen
  \bibfield  {author} {\bibinfo {author} {\bibfnamefont {V.~B.}\ \bibnamefont
  {Eltsov}}, \bibinfo {author} {\bibfnamefont {R.}~\bibnamefont {de~Graaf}},
  \bibinfo {author} {\bibfnamefont {R.}~\bibnamefont {H{\"a}nninen}}, \bibinfo
  {author} {\bibfnamefont {M.}~\bibnamefont {Krusius}}, \bibinfo {author}
  {\bibfnamefont {R.}~\bibnamefont {Solntsev}}, \bibinfo {author}
  {\bibfnamefont {V.~S.}\ \bibnamefont {L'vov}}, \bibinfo {author}
  {\bibfnamefont {A.~I.}\ \bibnamefont {Golov}}, \ and\ \bibinfo {author}
  {\bibfnamefont {P.~M.}\ \bibnamefont {Walmsley}},\ }in\ \href {\doibase DOI:
  10.1016/S0079-6417(08)00002-4} {\emph {\bibinfo {booktitle} {Progress in Low
  Temperature Physics: Quantum Turbulence}}},\ \bibinfo {series} {Progress in
  Low Temperature Physics}, Vol.~\bibinfo {volume} {16},\ \bibinfo {editor}
  {edited by\ \bibinfo {editor} {\bibfnamefont {B.}~\bibnamefont {Halperin}}\
  and\ \bibinfo {editor} {\bibfnamefont {M.}~\bibnamefont {Tsubota}}}\
  (\bibinfo  {publisher} {Elsevier},\ \bibinfo {year} {2009})\ pp.\ \bibinfo
  {pages} {45 -- 146}\BibitemShut {NoStop}%
\bibitem [{\citenamefont {Hosio}\ \emph {et~al.}(2011)\citenamefont {Hosio},
  \citenamefont {Eltsov}, \citenamefont {de~Graaf}, \citenamefont {Heikkinen},
  \citenamefont {H\"anninen}, \citenamefont {Krusius}, \citenamefont {L'vov},\
  and\ \citenamefont {Volovik}}]{Elt11}%
  \BibitemOpen
  \bibfield  {author} {\bibinfo {author} {\bibfnamefont {J.~J.}\ \bibnamefont
  {Hosio}}, \bibinfo {author} {\bibfnamefont {V.~B.}\ \bibnamefont {Eltsov}},
  \bibinfo {author} {\bibfnamefont {R.}~\bibnamefont {de~Graaf}}, \bibinfo
  {author} {\bibfnamefont {P.~J.}\ \bibnamefont {Heikkinen}}, \bibinfo {author}
  {\bibfnamefont {R.}~\bibnamefont {H\"anninen}}, \bibinfo {author}
  {\bibfnamefont {M.}~\bibnamefont {Krusius}}, \bibinfo {author} {\bibfnamefont
  {V.~S.}\ \bibnamefont {L'vov}}, \ and\ \bibinfo {author} {\bibfnamefont
  {G.~E.}\ \bibnamefont {Volovik}},\ }\href@noop {} {\bibfield  {journal}
  {\bibinfo  {journal} {Phys. Rev. Lett.}\ }\textbf {\bibinfo {volume} {107}},\
  \bibinfo {pages} {135302} (\bibinfo {year} {2011})}\BibitemShut {NoStop}%
\bibitem [{\citenamefont {Sonin}\ and\ \citenamefont {Nemirovskii}(2011)}]{SN}%
  \BibitemOpen
  \bibfield  {author} {\bibinfo {author} {\bibfnamefont {E.~B.}\ \bibnamefont
  {Sonin}}\ and\ \bibinfo {author} {\bibfnamefont {S.~K.}\ \bibnamefont
  {Nemirovskii}},\ }\href@noop {} {\bibfield  {journal} {\bibinfo  {journal}
  {Phys. Rev. B}\ }\textbf {\bibinfo {volume} {84}},\ \bibinfo {pages} {054506}
  (\bibinfo {year} {2011})}\BibitemShut {NoStop}%
\bibitem [{\citenamefont {Glaberson}\ \emph {et~al.}(1974)\citenamefont
  {Glaberson}, \citenamefont {Johnson},\ and\ \citenamefont
  {Ostermeier}}]{glabL}%
  \BibitemOpen
  \bibfield  {author} {\bibinfo {author} {\bibfnamefont {W.~I.}\ \bibnamefont
  {Glaberson}}, \bibinfo {author} {\bibfnamefont {W.~W.}\ \bibnamefont
  {Johnson}}, \ and\ \bibinfo {author} {\bibfnamefont {R.~M.}\ \bibnamefont
  {Ostermeier}},\ }\href@noop {} {\bibfield  {journal} {\bibinfo  {journal}
  {Phys. Rev. Lett.}\ }\textbf {\bibinfo {volume} {33}},\ \bibinfo {pages}
  {1197} (\bibinfo {year} {1974})}\BibitemShut {NoStop}%
\bibitem [{\citenamefont {Ostermeier}\ and\ \citenamefont
  {Glaberson}(1975)}]{glab}%
  \BibitemOpen
  \bibfield  {author} {\bibinfo {author} {\bibfnamefont {R.~M.}\ \bibnamefont
  {Ostermeier}}\ and\ \bibinfo {author} {\bibfnamefont {W.~I.}\ \bibnamefont
  {Glaberson}},\ }\href@noop {} {\bibfield  {journal} {\bibinfo  {journal} {J.
  Low Temp. Phys.}\ }\textbf {\bibinfo {volume} {21}},\ \bibinfo {pages} {191}
  (\bibinfo {year} {1975})}\BibitemShut {NoStop}%
\bibitem [{\citenamefont {Sonin}(1987)}]{RMP}%
  \BibitemOpen
  \bibfield  {author} {\bibinfo {author} {\bibfnamefont {E.~B.}\ \bibnamefont
  {Sonin}},\ }\href@noop {} {\bibfield  {journal} {\bibinfo  {journal} {Rev.
  Mod. Phys.}\ }\textbf {\bibinfo {volume} {59}},\ \bibinfo {pages} {87}
  (\bibinfo {year} {1987})}\BibitemShut {NoStop}%
\bibitem [{\citenamefont {Khalatnikov}(2000)}]{Khalatnikov2000}%
  \BibitemOpen
  \bibfield  {author} {\bibinfo {author} {\bibfnamefont {I.~M.}\ \bibnamefont
  {Khalatnikov}},\ }\href@noop {} {\emph {\bibinfo {title} {An Introduction to
  the Theory of Superfluidity}}}\ (\bibinfo  {publisher} {Perseus Publishing,
  Cambridge},\ \bibinfo {year} {2000})\BibitemShut {NoStop}%
\bibitem [{\citenamefont {{Raja Gopal}}(1964)}]{RG}%
  \BibitemOpen
  \bibfield  {author} {\bibinfo {author} {\bibfnamefont {E.~S.}\ \bibnamefont
  {{Raja Gopal}}},\ }\href@noop {} {\bibfield  {journal} {\bibinfo  {journal}
  {Ann. Phys. (N.Y.)}\ }\textbf {\bibinfo {volume} {29}},\ \bibinfo {pages}
  {350} (\bibinfo {year} {1964})}\BibitemShut {NoStop}%
\bibitem [{\citenamefont {Krusius}\ \emph {et~al.}(1993)\citenamefont
  {Krusius}, \citenamefont {Korhonen}, \citenamefont {Kondo},\ and\
  \citenamefont {Sonin}}]{ColModes}%
  \BibitemOpen
  \bibfield  {author} {\bibinfo {author} {\bibfnamefont {M.}~\bibnamefont
  {Krusius}}, \bibinfo {author} {\bibfnamefont {J.~S.}\ \bibnamefont
  {Korhonen}}, \bibinfo {author} {\bibfnamefont {Y.}~\bibnamefont {Kondo}}, \
  and\ \bibinfo {author} {\bibfnamefont {E.~B.}\ \bibnamefont {Sonin}},\
  }\href@noop {} {\bibfield  {journal} {\bibinfo  {journal} {Phys. Rev. B}\
  }\textbf {\bibinfo {volume} {47}},\ \bibinfo {pages} {15113} (\bibinfo {year}
  {1993})}\BibitemShut {NoStop}%
\bibitem [{\citenamefont {Bevan}\ \emph {et~al.}(1997)\citenamefont {Bevan},
  \citenamefont {Manninen}, \citenamefont {Cook}, \citenamefont {Alles},
  \citenamefont {Hook}, \citenamefont {Hook},\ and\ \citenamefont
  {Hall}}]{Bevan}%
  \BibitemOpen
  \bibfield  {author} {\bibinfo {author} {\bibfnamefont {T.~D.~C.}\
  \bibnamefont {Bevan}}, \bibinfo {author} {\bibfnamefont {A.~J.}\ \bibnamefont
  {Manninen}}, \bibinfo {author} {\bibfnamefont {J.~B.}\ \bibnamefont {Cook}},
  \bibinfo {author} {\bibfnamefont {H.}~\bibnamefont {Alles}}, \bibinfo
  {author} {\bibfnamefont {J.~R.}\ \bibnamefont {Hook}}, \bibinfo {author}
  {\bibfnamefont {H.~E. H. J.~R.}\ \bibnamefont {Hook}}, \ and\ \bibinfo
  {author} {\bibfnamefont {H.~E.}\ \bibnamefont {Hall}},\ }\href@noop {}
  {\bibfield  {journal} {\bibinfo  {journal} {J. Low Temp. Phys.}\ }\textbf
  {\bibinfo {volume} {109}},\ \bibinfo {pages} {423} (\bibinfo {year}
  {1997})}\BibitemShut {NoStop}%
\bibitem [{\citenamefont {Karim\"aki}\ \emph {et~al.}()\citenamefont
  {Karim\"aki}, \citenamefont {H\"anninen},\ and\ \citenamefont
  {Thuneberg}}]{KHT}%
  \BibitemOpen
  \bibfield  {author} {\bibinfo {author} {\bibfnamefont {J.~M.}\ \bibnamefont
  {Karim\"aki}}, \bibinfo {author} {\bibfnamefont {R.}~\bibnamefont
  {H\"anninen}}, \ and\ \bibinfo {author} {\bibfnamefont {E.~V.}\ \bibnamefont
  {Thuneberg}},\ }\href@noop {} {}\bibinfo {note} {{a}rXive:
  cond-mat/1108.5978.}\BibitemShut {Stop}%
\bibitem [{\citenamefont {Eltsov}()}]{EltPr}%
  \BibitemOpen
  \bibfield  {author} {\bibinfo {author} {\bibfnamefont {V.~B.}\ \bibnamefont
  {Eltsov}},\ }\href@noop {} {}\bibinfo {note} {(private communication,
  unpublished)}\BibitemShut {NoStop}%
\end{thebibliography}
\end{document}